\documentstyle[psfig,fleqn]{tp}

\begin{document}

\twocolumn[
\title{Velocity Fields from IRAS--PSC$z$ survey}
\author{E. Branchini$^1$, C.S. Frenk $^1$ 
L. Teodoro$^1$, I. Schmoldt$^2$, G. Efstathiou $^3$,\\
W. Saunders$^4$, S.D.M.  White$^5$,
M. Rowan--Robinson$^6$, \\
W. Sutherland$^2$,
H. Tadros$^2$, S. Maddox$^3$,
S. Oliver$^6$, O. Keeble$^6$ \\
{\it $^1$Department of Physics, University of Durham}\\
{\it $^2$Department of Physics, University of Oxford}\\
{\it $^3$Institute of Astronomy, Cambridge}\\
{\it $^4$Institute for Astronomy, University of Edinburgh}\\
{\it $^5$MPI--Astrophysik, Garching}\\
{\it $^6$Imperial College, University of London}}
\vspace*{16pt}   

ABSTRACT.

We present a self--consistent nonparametric model of the
cosmic velocity field based on the spatial distribution 
of IRAS galaxies in the recently completed 
all--sky PSC$z$ redshift survey.
The dense sampling of PSCz galaxies allows us to infer 
peculiar  velocities field up to large distances 
with unprecedented high resolution.
 
The most streaking feature of the PSCz model velocity field is 
a coherent large--scale streaming motion
along the  Perseus Pisces, Local Supercluster, Great Attractor 
and Shapley Concentration baseline, with no evidence for a backinfall 
into the Great Attractor region.
Instead, material behind and around the Great
Attractor in inferred  to be streaming towards the Shapley Concentration.

A likelihood analysis  that uses the information 
available on bulk velocities, cosmological dipoles and local 
shear
has been performed to measure $\beta$.
We have obtained $\beta =0.6^{+0.22}_{-0.15}$ (1--$\sigma$), 
in agreement with other recent determinations.

\endabstract]

\markboth{E. Branchini et al.}{PSC$z$ Velocity Field}

\small


\begin{figure*}
\centerline{\vbox{
\psfig{figure=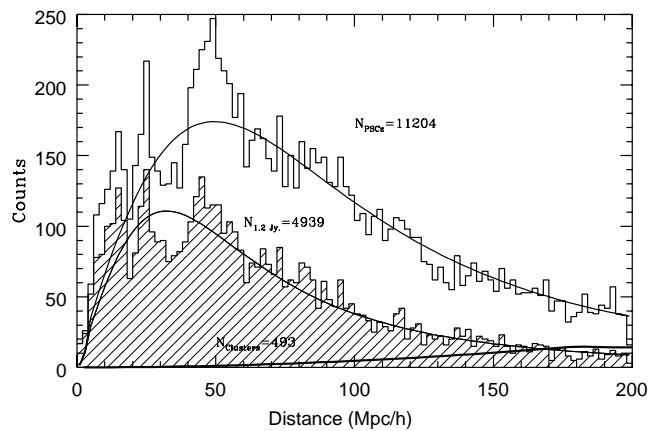,height=11cm}
}}
\vspace*{-2cm}
\caption[]{Distribution of IRAS PSC
galaxies vs. redshift. 
The upper and the lower, shaded histograms represent 
the  distribution of PSCz and 1.2Jy. galaxies, respectively.
The curves show the expected counts as a function of distance
estimated from the selection functions. 
The heavy line at the bottom shows
the predicted distance distribution of Abell/ACO clusters.
The labels give the total number of objects in each sample.}
\label{fig:1}
\end{figure*}

\section{Introduction}

The Gravitational Instability picture and
the Linear Biasing hypothesis provide us  
with a theoretical framework
in which galaxy peculiar velocities are related to the inhomogeneities
of the underlying mass density field, $\delta_m$.
On large scales, where Linear Theory applies, the relation 
is remarkably simple and one can compute peculiar velocities directly
from the galaxy distribution. 
The resulting velocity field is fully specified 
by the $\beta={{\Omega_m^{0.6}}\over{b}}$ parameter,  where  $b$ is the
bias parameter that linearly relates the inhomogeneities in the galaxy
distribution to the underlying mass density field.
 Comparing the model velocities with the observed ones
allows one to check the plausibility of the adopted theoretical scenario
and to constrain  the $\beta$ parameter.
Here we present a new model for the velocity field predicted 
from the PSC$z$ survey that we compare with existing peculiar 
velocity catalogues.

\section{The PSC$z$ Galaxy Survey}

Large sky coverage is the basic
requirement to  model  peculiar velocities from 
a redshift survey. 
The dataset used in this work is the recently completed PSC$z$
redshift survey described in detail by Saunders et al. (1998)
and briefly summarized here. 
The main
catalogue contains some 15,500 IRAS PSC galaxies with 60 $\mu m $ flux,
$f_{60}$, or greater.
To avoid cirrus contamination only PSC objects with $f_{100}<4 f_{60}$
were selected. Stars were excluded by requiring that $f_{60}>0.5 f_{25}$.
For our purposes, one of the most important properties of the PSC$z$
catalogue is its large sky coverage. The only excluded regions are two
thin strips in ecliptic longitude that were not observed by the IRAS
satellite, the Magellanic clouds, and the area in the galactic plane where
the B--band extinction exceeds 2 magnitudes.
Overall, the PSC$z$ catalogue covers $\sim 84 \%$ of the sky.
It is worth stressing that the survey is deeper and the 
galaxy sampling is denser that in any previous all--sky catalogue.
This point is clearly shown in Figure~1 in which the histogram
PSCz galaxy counts as a function of redshift is compared 
to the 1.2 Jy one (Fisher et al. 1995) and to the expected 
distribution inferred from the luminosity function (continuous lines).

\begin{figure*}
\centerline{\vbox{
\psfig{figure=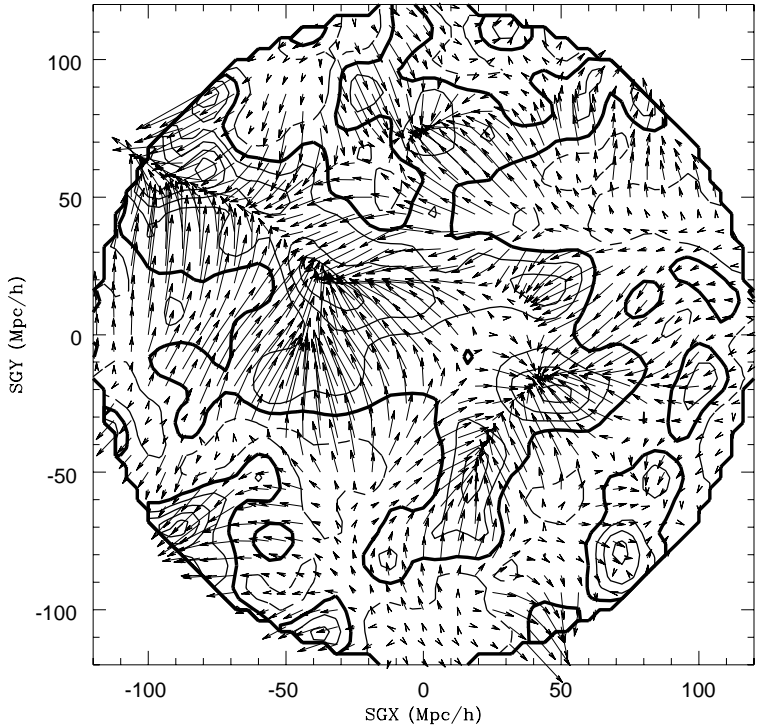,height=12.cm}
}}
\vspace*{-0.1cm}
\caption[]{PSC$z$ density and velocity fields within 120  $h^{-1}$ Mpc
on a slice through the Supergalactic plane. 
Continuous lines represent positive isodensity contours drawn
with a spacing of $\Delta \delta_m= 0.5$. 
The thick line sets the 0 overdensity level.
Negative contours are drawn with a dashed line.
The amplitude of the
velocity vectors, obtained for $\beta=1$, is on an arbitrary scale.
}
\label{fig:2}
\end{figure*}
 
\section{Building Velocity Models}

Systematic errors are a  major concern when modeling, measuring 
and comparing peculiar velocities. 
In the modeling phase we keep systematics under control
by using three independent techniques that predict
velocities from the galaxy distribution in redshift space:
 
\begin{itemize}
\item 1) A particle--based  iterative method similar to that 
introduced by Yahil et al (1991), which predict the peculiar 
velocity of the PSCz galaxies at their real space positions.
 
\item 2) A similar iterative algorithm in which, however,
densities and velocities are computed onto a regular grid.
 
\item 3) The Spherical Harmonics expansion technique introduced 
by Nusser and Davis (1994) in which the velocity field is predicted 
at any points in the redshift space.
 
\end{itemize}

All these methods assume linear or quasi--linear theory and therefore
require some degree of smoothing.

Random and systematic errors for the three model velocity fields  
have been evaluated using a suite of mock PSC$z$ catalogues 
extracted from the N--body simulations by Cole et al. (1997).
  
\begin{figure*}
\centerline{\vbox{
\psfig{figure=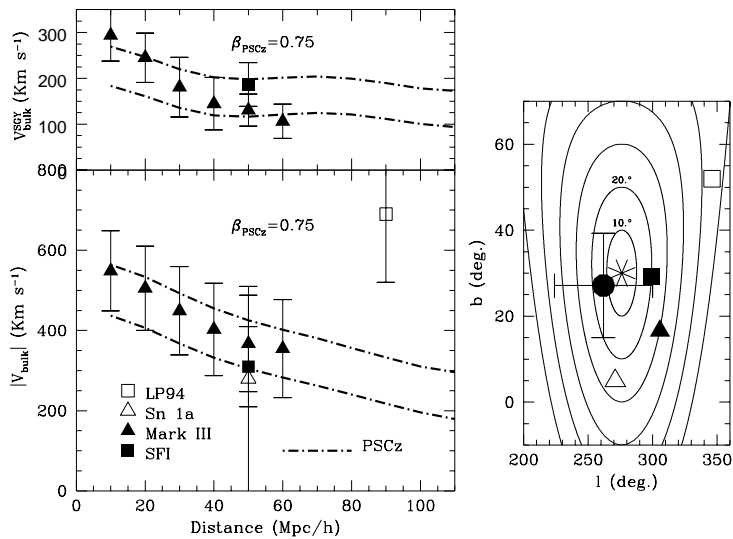,height=10.cm}
}}
\vspace*{-0.5cm}
\caption[]{
Model vs. observed bulk velocity.
Amplitude of the bulk flow vectors (lower left-hand) and
of its SGY--component (upper left-hand) are shown. 
The various symbols represent different experimental determinations. 
The dashed lines indicate the 1-$\sigma$ error strip from PSCz prediction.
The vectors' directions, measured at 50 $h^{-1}$ Mpc, are displayed
in the right-hand plot. The filled circle represents the prediction
from PSC$z$.
}
\label{fig:3}
\end{figure*}

\section{A Cosmographic Tour}

The depth and sampling frequency of the PSC$z$ dataset allow a reliable
map of the density and velocity fields 
to be constructed within a sphere around us of 120 $h^{-1}$ Mpc.
Figure~2 shows the PSC$z$ model density and velocity fields 
smoothed with a 6  $h^{-1}$ Mpc
Gaussian filter in a slice through the Supergalactic plane.

The Local Supercluster,
cluster at ${\rm (SGX,SGY)}=(-2.5,11.5)$, 
is connected to the prominent Hydra--Centaurus supercluster 
at ${\rm (SGX,SGY)}=(-35,20)$. Together with the
Pavo--Indus--Telescopium supercluster [${\rm (SGX,SGY)}=(-40,-15)$], the latter
makes up the well known Great Attractor. The Coma cluster 
appears as a peak at ${\rm (SGX,SGY)}=(0,75)$.
The Perseus--Pisces supercluster is
at ${\rm (SGX,SGY)}=(45,-20)$ and
its northern extension is visible at 
[${\rm (SGX,SGY)}=(45,20)$].
The   Cetus Wall may be seen as an elongated structure 
around ${\rm (SGX,SGY)}=(15,-50)$.
Finally, the Shapley Concentration starts appearing towards the edges at
[${\rm (SGX,SGY)}=(-90,60)$].
The Sculptor void 
is the largest underdense region at [${\rm (SGX,SGY)}=(-20,-45)$], 
almost matched in size by the void centered at
[${\rm (SGX,SGY)}=(70,50)$].

The local velocity field implied by the density field
is dominated by
the large infall patterns towards the Great Attractor, Perseus--Pisces and
Coma. A striking feature is the large-scale coherence of the velocity
field, apparent as a long ridge between Cetus and Perseus--Pisces and as a
large--scale flow along the Perseus--Pisces (North), Virgo, Great
Attractor, Shapley Concentration baseline. 
Note that a firm prediction of the PSC$z$ data 
is the lack of prominent back-infall onto the Great Attractor
region.

\section{The Bulk Flow}

The bulk velocity represents one of the 
simplest low--order statistics that,
in principle, can be estimated observationally and 
which has a theoretical counterpart. 
Measuring and modeling the bulk flow within a given volume,
however, is prone to random and systematic uncertainties.
Both errors have been estimated  using the mock PSC$z$ catalogues. 
In Figure~3 we compare our theoretical predictions 
to various observational measurements. 
On the left-hand side we show the amplitude of the bulk velocity 
(lower plot) and of its SGY component (upper plot).
The strip delimitated by the dashed lines represents the 
model prediction and its 1--$\sigma$ uncertainties.
Filled triangles are taken 
from Dekel et al. (1998) and represent the bulk flow from the Mark~III 
catalogue (Willick et al. 1997), 
measured using a POTENT--smoothing technique. A similar method
has been applied by Eldar et al. (1998) on the SFI catalogue
(Giovanelli et al. 1997). Their result is displayed with 
filled square. The open triangle shows the bulk flow obtained by considering 
the sample of SNe 1a (Riess et al. 1995). The open square displays
the result by Lauer and Postman (1994).
The right-hand side of the figure
shows the direction in the sky of the various bulk
flows. The filled circle represents the PSC$z$ prediction. The direction of 
the CMB dipole is plotted for reference and is represented by an
asterisk in the centre of the figure.

The bulk flow predicted using PSC$z$ agrees well, both in 
amplitude and directions,
with all the experimental determinations 
but, perhaps not surprisingly, disagrees with the result obtained by
Lauer and Postman. Comparing predicted and measured bulk flow 
allows one to measure $\beta$. As indicated in the figure,
we find $\beta \simeq 0.75$ with, however, a large error of
$\sim 50$ \%

\section{Estimating $\beta$}

Predicted peculiar velocities can be compared to the observed ones
to estimate $\beta$. This exercise, however, is potentially 
prone to systematic biases that need to be corrected for.
We have performed a likelihood analysis, similar to that 
introduced by Strauss et al. (1992), in which such a 
comparison can be performed that automatically accounts 
for random and systematic errors in a statistical fashion.
The observational quantities we consider are: the velocity of the 
Local Group measured from the CMB dipole
(${\bf v}_{LG}= 625 \pm 25$  km s$^{-1}$
from the 4 years COBE data of Linewaver et al. (1996)),
the bulk flow measured from the Mark III catalogue
by Dekel et al. (1998) and the 
evidence of a small ``shear'' ($|{\bf v}_s|<200$ km s$^{-1}$)
indicating that the local flow is remarkably cold (see van de 
Weygaert this volume). The theoretical quantities, predicted
from the distribution of PSC$z$ galaxies, are: 
the acceleration of at the position of the Local Group and 
the cumulative bulk flow in spheres of increasing radius.
Under the hypothesis that all the quantities mentioned above
are Gaussian random fields and assuming a prior model for the 
underlying cosmology it is possible to obtain an analytic
expression for their joint probability distribution.
Maximizing the distribution with respect to the free cosmological
parameters allows one to estimate  $\beta$.
We have obtained that $\beta= 0.6^{+0.22}_{-0.15}$ (1--$\sigma$).
Along with the constraints that this analysis sets on the other parameters,
we conclude that a universe with high $\Omega_m$ is less favourite by 
observations.

Other techniques of constraining $\beta$ involve point--by--point
velocity comparisons. Because of its dense sampling and low 
shot noise errors the PSC$z$ velocities can be reliably modeled out to 
large distances and then compared with objects capable of tracing
the large scale dynamics, such as clusters of galaxies.
A velocity--velocity comparison of this type 
has been recently performed using the PSC$z$ model velocity field and  the SMAC
catalogue of clusters' velocities and is described by Hudson et al. in
this volume.

\section*{Acknowledgments}

We thank Michael Strauss and Marc Davis for providing us
with the original version of their reconstruction codes
and Avishai Dekel 
for providing us with  unpublished bulk velocity data.

\end{document}